\documentclass{article}

\usepackage{PRIMEarxiv}

\usepackage[utf8]{inputenc} 
\usepackage[T1]{fontenc}    
\usepackage[colorlinks=true,
            linkcolor=blue,
            urlcolor=magenta,
            citecolor=green,
            anchorcolor=red]{hyperref}       
\usepackage{url}            
\usepackage{booktabs}       
\usepackage{amsfonts}       
\usepackage{bbold}
\usepackage{nicefrac}       
\usepackage{microtype}      
\usepackage{lipsum}
\usepackage{fancyhdr}       
\usepackage{graphicx}       

\usepackage{algorithm}
\usepackage{algpseudocode}
\usepackage{color}
\newcommand{\xhdr}[1]{\vspace{5pt} \noindent {\textbf{#1}}}

\usepackage[font=small,skip=3pt]{caption}
\usepackage{amsmath}
\usepackage{braket}
\makeatletter
\DeclareRobustCommand\onedot{\futurelet\@let@token\@onedot}
\def\@onedot{\ifx\@let@token.\else.\null\fi\xspace}

\DeclareMathOperator{\tr}{tr}

\title{Scalable neural quantum states architecture for quantum chemistry
}

\author{
Tianchen Zhao$^{1}$, \,James Stokes$^{2}$, 
\,Shravan Veerapaneni$^{1,2}$ \vspace{0.05in}\\ 
  $^{1}$Department of Mathematics, University of Michigan, Ann Arbor, MI 48109\vspace{0.05in} \\
  $^{2}$Flatiron Institute, Simons Foundation, New York, NY 10010 \vspace{0.05in}\\
  \texttt{\{ericolon, shravan\}@umich.edu}\\
  \texttt{jstokes@flatironinstitute.org}\\
}

\begin{document}
\maketitle

\begin{abstract}
Variational optimization of neural-network representations of quantum states has been successfully applied to solve interacting fermionic problems. Despite rapid developments, significant scalability challenges arise when considering molecules of large scale, which correspond to non-locally interacting quantum spin Hamiltonians consisting of sums of thousands or even millions of Pauli operators. In this work, we introduce scalable parallelization strategies to improve neural-network-based variational quantum Monte Carlo calculations for ab-initio quantum chemistry applications. We establish GPU-supported local energy parallelism to compute the optimization objective for Hamiltonians of potentially complex molecules. Using autoregressive sampling techniques, we demonstrate systematic improvement in wall-clock timings required to achieve CCSD baseline target energies. The performance is further enhanced by accommodating the structure of resultant spin Hamiltonians into the autoregressive sampling ordering. The algorithm achieves promising performance in comparison with the classical approximate methods and exhibits both running time and scalability advantages over existing neural-network based methods.
\end{abstract}


\section{Introduction}


Quantum many-body systems describe a vast category of physical problems at microscopic scales. In the context of ab-initio Quantum Chemistry (QC), the central topic is to unravel the quantum effects determining the structure and properties of molecules by solving the many-body time-independent Schr\"odinger equation describing the interaction of heavy nuclei with orbiting electrons. However, the exponential complexity with respect to the number of particles makes the analytical computations about the system impractical~\cite{troyer-pr05}.

Classical strategies discretize the problem using a finite number of basis functions, expanding the full many-body state in a basis of anti-symmetric Slater determinants. Because of the exponential scaling of the determinant space, however, exact approaches that systematically consider all electronic configurations such as the full configuration interaction (FCI) method, are typically restricted to small molecules and basis sets. Coupled cluster (CC) techniques~\cite{coester-60short,bartlett-07coupled} are approximate methods routinely adopted in QC electronic calculations, however, the accuracy of CC is intrinsically limited in the presence of strong quantum correlations, in turn restricting the applicability of the method to regimes of relative weak correlations.
Recently neural-network quantum state (NQS) methods~\cite{choo-20fermionic,barrett2022autoregressive} have proven to be successful variational ansatze for finding the ground state of molecular systems up to 30 qubits ($\text{Li}_2\text{O}$). However, significant scalability challenges of the NQS approach arise when considering molecules of larger scales. The scalability issue stems from two sources, which we refer to as \emph{local energy parallelism} and \emph{sampling parallelism}. (\emph{i}) The complexity of the computation of local energy scales linearly with respect to the number of terms in the molecular Hamiltonian, which induces out-of-memory (OOM) issues for larger problems. (\emph{ii})
In order to achieve satisfactory performance, one needs to sample exact and accurate configurations from the targeted distribution, which becomes expensive or even impossible for existing approaches such as Markov chain Monte-Carlo (MCMC) in high dimensions.

In order to address local energy parallelism for large-scale molecules, we use an efficient tensor representation of the second quantized spin Hamiltonian generated from chemical data so that the computation of the local energy is efficiently supported by GPUs. We further employ identical copies of the model across the computing units to generate only a few samples per unit and combine the independent samples from all these units to construct an accurate expectation estimate. In addition, memory consumption is reduced through gradient accumulation, which splits the batch of samples used for training the model into several mini-batches of samples that  
can be distributed across the devices and processed independently.
The proposed parallelization scheme for large batch computation is particularly important for chemical molecules with a large number of terms in their electronic Hamiltonian formulation, where the calculation of local energy becomes prohibitive.

The basis of our approach to achieving sampling parallelism is our utilization of a wavefunction based on Masked Autoencoder for Distribution Estimation (MADE)~\cite{germain-icml15}. Using MADE as our base model enables the development of a parallelization scheme based on \cite{zhao2021overcoming}. Unlike restricted Boltzmann machines (RBMs) \cite{salakhutdinov2008quantitative} and NADE~\cite{larochelle2011neural}, which have formed the basis of previous ab-initio studies, MADE is known to be lightweight and scalable to high-dimensional inputs. On the one hand, it overcomes the asymptotical convergence issues of MCMC by using autoregressive sampling. On the other hand, it bypasses the inherently sequential nature of NADE with the minor additional cost of simple masking operations.  In addition, we further improve the performance of our model through the autoregressive sampling of the state entries in an order that roughly matches the entanglement hierarchy among the involved qubits. Our experiments demonstrate that the proposed algorithm effectively works for molecules up to 76 qubits with millions of terms in its electronic Hamiltonian.

The paper is organized as follows. In section \ref{sec:related}, we begin by summarizing the existing state-of-art in neural-network quantum state research as applied to ab initio quantum chemistry. Section \ref{sec:formulation} provides mathematical details about the nature of the Hamiltonians under consideration as well as stochastic approximation strategy based on neural networks. The proposed neural-network quantum state architecture is elaborated in section \ref{sec:modeling} and the parallel evaluation strategies are subsequently detailed in section \ref{sec:parallel}. Experimental results for molecules up to 76 qubits are described in section \ref{sec:experiments}, including ablative studies on the relevant factors controlling the performance of the algorithm.

\section{Related Work}\label{sec:related}

\xhdr{Variational Monte-Carlo and Autoregressive Quantum States.} The idea of utilizing neural-network quantum states to overcome the curse of dimensionality in high-dimensional VQMC simulations was first introduced by Carleo and Troyer \cite{carleo2017solving}, who concentrated on restricted Boltzmann machines (RBMs) applied to two-dimensional quantum spin models. However, RBM relies on approximate sampling procedures like MCMC, whose convergence time remains undetermined, which often results in the generation of highly correlated samples and deterioration in performance. Such sampling approximations can be avoided by using autoregressive models~\cite{bengio2000modeling} that estimate the joint distribution by decomposing it into a product of conditionals by the probability chain rule, making both the density estimation and generation process tractable. To this end, Larochelle and Murray~\cite{larochelle2011neural} proposed neural autoregressive distribution estimator (NADE) as feed-forward architectures. MADE~\cite{germain-icml15} improves the efficiency of models with minor additional costs for simple masking operations. Sharir \emph{et al.} \cite{sharir-prl20,sharir-git20} and Hibat-Allah \cite{hibat2020recurrent} introduced neural-network quantum states based on the autoregressive assumption inspired, respectively by PixelCNN \cite{oord2016pixelcnn} and recurrent neural networks, and demonstrated significantly improved performance compared to RBMs.

\xhdr{Application in Quantum Chemistry.} Many approximate methods specific to the QC problem~\cite{hammond-94monte,langhoff-12quantum} have been discovered by researchers. The Hartree-Fock approximation treats each electron in the molecule as an independent particle that moves under the influence of the Coulomb potential due to the nuclei, and a mean-field generated by all other electrons. It calculates the expansion coefficients of the linear combination of atomic orbitals.
Configuration interaction methods~\cite{sherrill-99advances} use a linear combination of configuration state functions built from spin orbitals to restrict the active space to configuration strings.
Coupled cluster approaches~\cite{coester-60short,bartlett-07coupled} construct multi-electron wavefunctions using the exponential cluster operator to account for electron correlation, but cannot parameterize arbitrary superpositions and occasionally lead to unphysical solutions. Choo \emph{et al.}~\cite{choo-20fermionic} proposed an RBM-based NQS  variational ansatz, leveraging the power of artificial neural networks that have recently emerged in the more general context of interacting many-body quantum matter \cite{carleo2017solving}. This approach provides a compact, variational parameterization of the many-body wave function. Barrett \emph{et al.}~\cite{barrett2022autoregressive} subsequently proposed a novel autoregressive NQS architecture for ab-initio QC based on NADE \cite{larochelle2011neural} with hard-coded pre-and post-processing that enables exact sampling of physically viable states. These advances ultimately allow their approach to scale to systems up to 30 qubits, surpassing what was previously possible using RBMs, while still falling short of large-scale applications. Empirically, we have found that the approach of \cite{barrett2022autoregressive} encounters significant scalability challenges in generalizing to molecular systems of yet larger size.
\section{Problem Formulation}
\label{sec:formulation}

We will work under the assumption of the so-called {\em Born-Oppenheimer approximation} \cite{born1927quantentheorie}, which treats the nuclei as fixed point charges. Indeed, the specification of a molecular geometry implicitly assumes such an approximation. When the positions of the nuclei are specified, the electronic structure problem can be restated as finding the ground eigenstate of the electronic Hamiltonian operator and consequently the ground state electronic energy becomes a parametric function of atomic positions. Here, the molecule’s electronic Hamiltonian is commonly represented using the second-quantization formalism, with a chosen basis of atomic orbitals which describe the wave function of electrons in the molecule. In order to identify the Hamiltonian for a compound, we start by fetching the information required to build the target molecule object, then employ established solvers to compute the second-quantized Hamiltonian and store the result in a string format.

In the second-quantized formulation, the Hamiltonian is represented as a complex conjugate-symmetric (Hermitian) matrix $H$ of side length $2^n$ where $n$ represents the number of orbitals, and the basic problem is to determine the ground energy of $H$ and a description of an associated eigenvector. Since storage and manipulation of such matrices is prohibitive, a common practice is to exploit the fact that $H$ admits an efficient description in terms of superpositions of tensor products of $2\times 2$ matrices drawn from the set $\{ \sigma_0, \sigma_1,\sigma_2,\sigma_3 \}$, where
\begin{equation}
    \sigma_0
    =
    \begin{bmatrix}
    1 & 0 \\
    0 & 1
    \end{bmatrix}
    \enspace , \quad \quad
    \sigma_1
    =
    \begin{bmatrix}
    0 & 1 \\
    1 & 0
    \end{bmatrix}
    \enspace , \quad \quad
    \sigma_2
    =
    \begin{bmatrix}
    0 & -i \\
    i & 0
    \end{bmatrix}
    \enspace , \quad \quad
    \sigma_3
    =
    \begin{bmatrix}
    1 & 0 \\
    0 & -1
    \end{bmatrix}
\end{equation}

In fact, any Hermitian matrix $H \in \mathbb{C}^{2^n \times 2^n}$ admits a unique decomposition of the form
\begin{equation}\label{e:hamiltonpauli}
    H = \sum_{\mathbf{p} \in \{0,1,2,3\}^n} \alpha_{\mathbf{p}} \, P_{\mathbf{p}}
\end{equation}
where $P_\mathbf{p} := \sigma_{p_1} \otimes \cdots \otimes \sigma_{p_n}$ denotes a tensor product of Pauli matrices, and the real-valued coefficients entering the sum are determined by the formula $\alpha_\mathbf{p} = \frac{1}{2^n}\tr(H P_\mathbf{p})$. Since each matrix $P_\mathbf{p}$ is row-sparse with exactly one nonzero entry per row, the number of nonzero entries per row of $H$ is bounded by the number of nonzero coefficients $K$ entering the above sum, which could be as large as $4^n$ for typical matrices.
The matrices of relevance to quantum chemistry are far from typical, however, and consequently enjoy a high level of row sparsity, which will be a crucial property in our subsequent algorithm development.

Beyond the assumption of row-sparsity, additional features of $H$ can be determined from the structure of the constituent Pauli strings $\mathbf{p}$, which in turn depend on the choice of encoding map from the chemical Hamiltonian to the qubit representation. Efficient qubit encoding maps are an active field of research, and the most common examples in usage are Jordan-Wigner \cite{jordan-93} and Bravyi-Kitaev \cite{bravyi-02fermionic}, which are equivalent isospectrally. In this work, we will adopt the Jordan-Wigner encoding because of its simplicity and the fact that it has shown success in prior work on VMC applied to ab-initio quantum chemistry \cite{choo-20fermionic,barrett2022autoregressive}. In the particular case of Jordan-Wigner transformation, we point out that many of the tensor factors of terms appearing in the sum \eqref{e:hamiltonpauli} are the identity matrix and are moreover organized in a hierarchical structure.

Following the standard neural-network variational Monte Carlo procedure, we postulate a family of trial wavefunctions whose complex amplitudes relative to the standard basis are computed by the output of a neural network, parametrized by variational parameters $\theta \in \mathbb{R}^d$. Thus, given a function of the form
\begin{equation}
    f : \{0,1\}^n \times \mathbb{R}^d  \longrightarrow \mathbb{C}
\end{equation}
which is differentiable in the second argument, we define an associated family of trial quantum states $|f_\theta\rangle$, which are differentiably parametrized by $\theta \in \mathbb{R}^d$, as follows
\begin{equation}\label{e:trial}
    | f_\theta \rangle := \sum_{x \in \{0,1\}^n} f(x,\theta) | x \rangle
\end{equation}
where $|x\rangle := |x_1\rangle \otimes \cdots \otimes |x_n\rangle$ is a shorthand to denote the standard basis vectors for $\mathbb{C}^{2^n}$. In this work, following \cite{barrett2022autoregressive}, we assume that the function $f$ is chosen in such a way that $|f_\theta\rangle$ is unit-normalized for all $\theta \in \mathbb{R}^d$, and that exact sampling from the probability distribution $\pi_\theta := |f(\cdot,\theta)|^2$ is computationally tractable. Using the unit vector \eqref{e:trial} as a trial vector in the Rayleigh quotient for $H$, we obtain a differentiable objective function $\mathcal{L} : \mathbb{R}^d \longrightarrow \mathbb{R}$, which upper bounds the minimal eigenvalue $\lambda_{\min}(H)$ as a consequence of the Rayleigh-Ritz principle,
\begin{equation}
    \mathcal{L}(\theta) := 
    \frac{1}{2} \langle f_\theta | H | f_\theta \rangle
\end{equation}
and whose value can be estimated at the Monte Carlo rate using the following estimator
\begin{equation}\label{eq:total_energy}
    \mathcal{L}(\theta) = \frac{1}{2} \mathbb{E}[l_\theta(x)] \enspace , \quad \quad l_\theta(x) := \frac{\langle x | H | f_\theta \rangle}{f(x,\theta)} \enspace , \quad \quad x \sim \pi_\theta
\end{equation}
where $l_\theta$ is referred to as the local energy.
Minimization of $\mathcal{L}$ is performed using stochastic gradient-based optimization techniques. In particular, the gradient of $\mathcal{L}$ can be estimated at the Monte Carlo rate using,
\begin{equation}\label{e:grad}
    \nabla \mathcal{L}(\theta) = \operatorname{Re} \mathbb{E}
    \left[ \left(l_\theta(x)\mathbb{1} - b \right) \overline{\sigma_{\theta}}(x) \right] \enspace ,
    \quad \quad \sigma_\theta(x) := \frac{\nabla_\theta f(x,\theta)}{f(x,\theta)} \enspace ,
    \quad \quad x \sim \pi_\theta
\end{equation}
where $b \in \mathbb{R}^{d\times d}$ is an arbitrary baseline matrix that can be set to $b=\mathbb{E}[l_\theta(x)]$.

Eq.~\eqref{e:grad} demonstrates that an integral component of the algorithm is the computation of the local energy, which is required both to perform stochastic gradient updates of the model and to estimate the value of the  objective function. The computational complexity of computing the mapping $x \longmapsto l_\theta(x)$ for a minibatch of size $B$ is evidently $O(BK)$ where recall that $K$ denotes the number of terms in the Hamiltonian expansion \eqref{e:hamiltonpauli}. Despite the fact that the sparsity parameter satisfies $K \ll 4^n$, the computation of local energy still suffers from severe OOM issues in practice since $4^n$ can be extraordinarly large even for modestly sized molecules. For example, in the case of Methanol with $n = 28$ orbitals we have $4^n \approx 7.2 \times 10^{16}$ while $K = 52887$. This implies that for Methanol, given a modest batch size $B=1024$, the local energy needs a forward pass of $54$M samples of input size $28$.
This computational bottleneck is inevitable as large batch sizes are essential to ensure the precision of the Monte-Carlo approximation of the energy objective in Eq.~\eqref{eq:total_energy}, which directly influences the performance of the algorithm.
Therefore, efficient sampling and local energy computation become the key issue for scaling the neural-network variational approach to large compounds.

\section{Autoregressive Modeling of Molecular Quantum States}\label{sec:modeling}

Motivated by the fact that solving high dimensional molecular quantum systems is still a difficult problem with existing variational techniques, we consider the base model from Zhao \emph{et al.}~\cite{zhao2021overcoming}, which was shown to be capable of solving quantum systems of very high dimensions. Adaptation of this work to chemistry problems involves generalizing from real-valued to complex-valued wave functions, as well as adjusting the sampling process to handle larger numbers of configurations and accommodate domain priors. In addition, following \cite{barrett2022autoregressive}, we enforce constraints necessary to ensure that the generated samples correspond to physical electronic states.

\xhdr{Architecture.}
Since the ground state of the targeted problem is in general complex-valued, we consider the complex generalization of the model in \cite{zhao2021overcoming} by splitting the output of the wave function into modulus and phase parts, learned by two sub-models separately as follows
\begin{align}
    \text{Modulus sub-model: } \textit{Input} &\xrightarrow{[B,N]} \texttt{MaskedFC1} \xrightarrow{[B,h]} \texttt{ReLU} \nonumber \\ &\xrightarrow{[B,h]} \texttt{MaskedFC2} \xrightarrow{[B,N]} \texttt{Sigmoid} \xrightarrow{[B,N]} \textit{Output}, \nonumber \\
    \text{Phase sub-model: } \textit{Input} &\xrightarrow{[B,N-2]} \texttt{MaskedFC1} \xrightarrow{[B,h]} \texttt{ReLU} \xrightarrow{[B,h]} \texttt{MaskedFC2} \xrightarrow{[B,4]} \textit{Output}. \nonumber
\end{align}
The modulus model predicts $n$ conditional probabilities for each configuration, and the phase model predicts the phases for each of the four configurations that are identical in the first $n-2$ entries, which is a $n-2$ dimensional vector being fed as the input.
Here $B$ is the batch size, $n$ is the number of dimensions, $h$ is the hidden layer size and $\texttt{MaskedFC}$ is the masked fully connected layer, which removes the connections in the computational path of MADE. The outputs of modulus sub-model are the conditional probabilities $\{ \pi_i(x_i|x_{i-1},\ldots,x_1) \}_{i=1}^n$, which together define a joint probability function $\pi : \{0,1\}^n \longrightarrow [0,\infty)$ for input strings $x$. Normalization follows automatically from the autoregressive assumption,
\begin{equation}
    \pi(x) = \prod_{i=1}^n \pi_i(x_i|x_{i-1},\ldots,x_1) \enspace .
\end{equation}
On the other hand, we model the phase $\phi : \{0,1\}^n \longrightarrow[0,2\pi)$ of the wavefunction directly by feeding the input $x$ into a two-layer MLP. It follows that the complex logarithm of the model output $f(x,\theta)$ can be written in a computationally tractable form as
\begin{align}
    f(x,\theta) &= e^{i\phi(x)} \prod_{i=1}^n \sqrt{\pi_i(x_i|x_{i-1},...,x_1)} , \\
    \log f(x,\theta) &= i\phi(x) + \frac{1}{2}\sum_{i=1}^n \log \pi_i(x_i|x_{i-1},...,x_1)
\end{align}
where $\theta \in \mathbb{R}^d$ denotes the concatenation of parameters describing the neural networks $\phi$ and $\pi$. 

\xhdr{Sampling.}
Standard autoregressive sampling techniques such as NADE \cite{larochelle2011neural} have a sequential nature that updates a batch of randomly initialized states entry by entry following a pre-fixed order. In practice, this approach is highly inefficient as the sampled batch usually contains repeated samples. Instead, we keep track of a sample buffer throughout the sampling process, associated with a counter storing the number of occurrences for each sample in the buffer. We start with a buffer containing only one random initialized sample. At each of the $n$ iterations, we first double the size of the buffer by alternating $\pm 1$ value for existing samples at a fixed entry. Then, we update the counter for all samples in the buffer through Bernoulli sampling with probabilities computed by forward pass. Finally, we eliminate the samples in the batch that have the lowest numbers of occurrences in the counter, to avoid the exponential growth of the buffer size.  

In addition, as a consequence of the Jordan-Wigner encoding, many of the tensor factors appearing with high qubit index are given by the identity factor, which leads to  the expectation that high index qubits are comparatively less correlated compared to those with at low indices. Motivated by this observation, in an effort to ease the training of the model we performed autoregressive sampling in a reversed order beginning with the $n$th qubit.

\begin{figure}[t]
\centering
\includegraphics[width=0.95\linewidth]{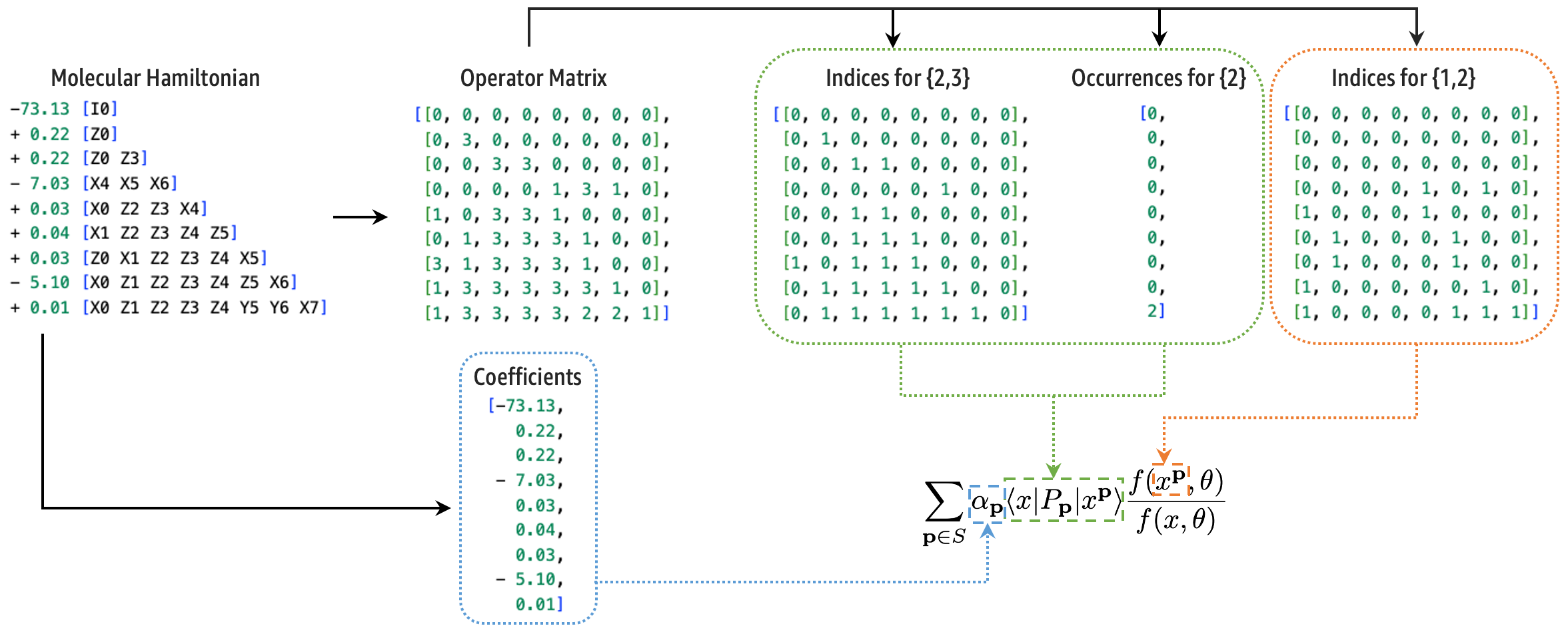}
\caption{Local energy computational paradigm. The Hamiltonian is parsed into an operator matrix and a coefficient vector, from which incidence matrices $F^{\{1,2\}},F^{\{2,3\}} \in \{0,1 \}^{K \times n}$ are extracted for subsequent calculation of the local energy. This formulation provides significant computational speed-up for molecular Hamiltonian with a huge number of terms. \label{fig:local_energy}}
\end{figure}

\xhdr{Constraints.}
Recall that the unconstrained model assigns nonzero probability mass to all $2^n$ possible bit-strings representing possible states of a multi-electron system. However, in quantum chemistry problems we considered, only $n_e \leq n$ out of the $n$ single-electron spin-orbitals are occupied with electrons, and the net charge $C$ of the molecular system determines the number of unpaired electrons. It follows that the numbers of electrons with up-spins and down-spins $n_\uparrow ,n_\downarrow$ respectively should satisfy
\begin{align}
    n_\uparrow +n_\downarrow=n_e,\quad n_\uparrow -n_\downarrow=C.
\end{align}
This corresponds to applying Hamming weight constraints to the bit-strings, which effectively reduces the total number of candidate samples from $2^{n}$ to ${{n/2}\choose{n_\uparrow }}{{n/2}\choose{n_\downarrow}}$, which significantly reduces the complexity of the problem. We adopted techniques from similar work~\cite{hibat2020recurrent,barrett2022autoregressive} to enforce the constraints. Define the hamming weight constraint of the configuration $x \in \{0,1\}^n$ to be
\begin{align}
    \sum_{i=1}^{n/2} x_{2i-1} = n_\uparrow , \quad \sum_{i=1}^{n/2} x_{2i}=n_\downarrow,
    \label{eq:hamming_weight_constraint}
\end{align}
where the even and odd indices correspond to the up-spin and down-spin electrons in the orbitals.
The idea to enforce the constraint in the autoregressive sampling process is to assign nonzero probabilities only to samples that satisfy Eq.~\eqref{eq:hamming_weight_constraint}, for which the sufficient and necessary condition is that the $k$th entry $x_k$ must satisfy
\begin{align}
    n_\uparrow  - (n/2 - \lceil k/2 \rceil) \leq \sum_{i=1}^{\lceil k/2 \rceil} x_{2i-1} \leq n_\uparrow , \quad n_\downarrow - (n/2 - \lceil k/2 \rceil) \leq \sum_{i=1}^{\lceil k/2 \rceil} x_{2i} \leq n_\downarrow,
    \label{eq:hamming_weight_constraint_condition}
\end{align}
for all $k \in {1,2,...,N}$. Condition~\eqref{eq:hamming_weight_constraint_condition} can be enforced at every iteration $k$ during the sampling process by introducing the following modified probability distribution,
\begin{align}
    \widehat{\pi}(x_k=1|x_{k-1},...,x_1)=
    \begin{cases}
        1                                       , & \text{if } \sum_{i=1}^{\lceil k/2 \rceil} x_{2i-1} < n_\uparrow  - (n/2 - \lceil k/2 \rceil) \\
        1                                       , & \text{if } \sum_{i=1}^{\lceil k/2 \rceil} x_{2i} < n_\downarrow - (n/2 - \lceil k/2 \rceil) \\
        0                                       , & \text{if } \sum_{i=1}^{\lceil k/2 \rceil} x_{2i-1} \geq n_\uparrow  \\
        0                                       , & \text{if } \sum_{i=1}^{\lceil k/2 \rceil} x_{2i} \geq n_\downarrow \\
        \pi(x_k=1|x_{k-1},...,x_1) , & \textit{o.w.}, \\
    \end{cases}
\label{eq:modifier}
\end{align}
which has the property that for any $x$ violating the constraints, we have $\widehat\pi(x_k=x|x_{k-1},...,x_1) =0$ and $\widehat\pi(x_k=1-x|x_{k-1},...,x_1)=1$. The modified probabilities are still normalize to one by design.

\section{Parallelization}\label{sec:parallel}

We parallelize the training by distributing the input batch across the GPUs, where the model parameters are replicated on each GPU, which handles a portion of the full batch. During the backward pass, gradients from each node are averaged. Locally within each process, we tensorize the entire computational pipeline so that the computation of local energies is fully GPU-supported.

\begin{algorithm}[t]
\caption{Parallel Tensor Computation of Local Energy (presented for batch size 1)}
\begin{algorithmic}
    \State \textbf{Input}: Bit string $x \in \{0,1 \}^n$, coefficient vector, incidence matrices $F^{\{1,2\}},F^{\{2,3\}} \in \{0,1 \}^{K \times n}$
    \State \textbf{Output}: Local energy
    \State Compute $\{x^\mathbf{p}\} \in \{0,1\}^{K \times n}$ as $\{x^\mathbf{p}\}=X\oplus F^{\{1,2\}}$, where $X\in\{0,1\}^{K\times n}$ is the $K$-fold replication of $x$
    \State  Compute amplitudes $f(x)\in \mathbb{C}$ and $\{f(x^\mathbf{p})\} \in \mathbb{C}^K$
    \State  Compute $\{\langle x | P_\mathbf{p}|x^\mathbf{p}\} \in \mathbb{C}^K$ using Eq.~\eqref{eq:mtx_element}
    \State Evaluate the sum in Eq.~\eqref{eq:local_energy_auto}
\end{algorithmic}
\label{algo:automatic_local_energy}
\end{algorithm}


\subsection{Tensorized Computation of Local Energy}
The form of the local energy has the property that it can easily be parallelized, which becomes increasingly important with increasing molecular system size since the Hamiltonian can potentially involve a large number of terms. We implemented an efficient tensor representation of the second quantized spin Hamiltonian generated from chemical data, which capitalizes on the fact that the matrix corresponding to an arbitrary product of Pauli operators $P_{\mathbf{p}}$ is extremely sparse. In particular, for each row index $x \in \{0,1\}^n$ there is exactly one column index $x^\mathbf{p} \in \{0, 1\}^n$, for which the corresponding matrix entry $\langle x | P_{\mathbf{p}} | x^\mathbf{p} \rangle$ is nonzero\footnote{The formula for the associated matrix entry in terms of $x$ is given in \cite{choo-20fermionic}.}. If we denote the subset of Pauli strings with nonzero coefficient by $S = \{ \mathbf{p} \in \{0,1,2,3\}^n : \alpha_\mathbf{p} \neq 0 \}$, then the local energy simplifies to
\begin{align}
l_\theta(x) = \sum_{\mathbf{p} \in S} \alpha_{\mathbf{p}} \langle x | P_{\mathbf{p}} | x^\mathbf{p} \rangle \frac{f(x^\mathbf{p},\theta)}{f(x,\theta)} \enspace .
\label{eq:local_energy_auto}
\end{align}
The goal now is to efficiently compute Eq.~\eqref{eq:local_energy_auto}, in which the number of summands $K = |S|$ is very large. The idea, depicted in Fig.~\ref{fig:local_energy}, is to extract the key information required to perform the computation from the molecular Hamiltonian and store the information as tensors that directly support GPU computation. To this end, we constructed a string parser that computes an 
Operator Matrix along with coefficients. For Pauli string, we track the indices of Pauli $\sigma_1,\sigma_2,\sigma_3$ operators as tensors, which are later utilized to compute $x^\mathbf{p}$ and the corresponding matrix element $\langle x | P_{\mathbf{p}} | x^\mathbf{p} \rangle$.

In practice, $x^\mathbf{p}$ is computed by flipping the bits of $x$ based corresponding to the locations of $\sigma_1$ and $\sigma_2$ in $P_\mathbf{p}$. To improve the efficiency, we collect the indices of $\sigma_1,\sigma_2$ operators prior to the training and only keep a buffer of unique flippings and their corresponding number of occurrences. This approach can effectively reduce the input size to the forward pass by removing the repeated samples, which is particularly helpful for large-sized problems. The matrix element $\langle x | P_{\mathbf{p}} | x^\mathbf{p} \rangle$ admits the formula
\begin{align}
    \langle x | P_{\mathbf{p}} | x^\mathbf{p} \rangle = \braket{x|\sigma_{p_1} \otimes \cdots \otimes \sigma_{p_n}|x^\mathbf{p}}=(-i)^{r}\prod_{k:p_k\in\{2,3\}}(-1)^{x_k}
\label{eq:mtx_element}
\end{align}
where $r$ is the number of occurrences for the $\sigma_2$ operator, \textit{i.e.}, $r=\sum_{k=1}^n \mathbb{1}_{\{p_k=2\}}$. Similar to before, the indices of $\sigma_2, \sigma_3$ operators are collected prior to the training, and the product can be calculated by the Hadamard product between the indices and $x$, followed by the production of all entries. Note that all computations in this subsection can be performed in parallel for all terms with GPU.

\vspace{-1mm}

\subsection{Parallel algorithm implementation}

In applications such as quantum chemistry, the Hamiltonian of even the smallest molecules contain thousands of terms, which leads to severe OOM issues for the existing VMC platforms. Our proposed pipeline tensorizes the information in the molecular Hamiltonian to maximize memory efficiency. In addition, the computation of local energy is conducted term-wise with no interaction between the terms, which motivates embarrassingly parallel algorithms for Hamiltonians consisting of a large number of terms. We take a step toward addressing the bottleneck by applying our sampling parallelization strategy to this problem, where we use identical copies of the model across the computing units to generate only a few samples per unit and combine the independent samples from all these units to construct an accurate expectation estimate. In addition, we remove the replicated configurations as described in the previous section before the forward pass locally for each GPU to save more memory.

In more detail, recall that the energy expectation is approximated as the following double sum
\begin{align}
    \frac{\braket{f_{\theta}|H|f_{\theta}}}{\braket{f_{\theta}|f_{\theta}}}
    \approx \frac{1}{B}\sum_{i=1}^B\sum_{\mathbf{p} \in S} \alpha_{\mathbf{p}} \langle x_i | P_{\mathbf{p}} | x_i^\mathbf{p} \rangle \frac{f(x_i^\mathbf{p},\theta)}{f(x_i,\theta)} \enspace .
\label{eq:total_energy_sum}
\end{align}
We distribute the $BK$ summands in the above sum across $L$ GPUs in order to perform the forward and backward passes of the model with a mini-batch size of $BK/L$. Locally, each GPU has access to the necessary ingredients required to compute the corresponding partial sums of size $BK/L$. We compute local gradients with forward and backward passes within each GPU and update the model parameters with the averages of local gradients obtained by averaging over $BK/L$ elements. In addition, we use gradient accumulation \cite{lin2017deep}, splitting the batch into several mini-batches before a single update to avoid potential OOM issues for molecules with larger sizes.

\begin{table}[t]
\scriptsize
\centering
\begin{tabular}{lccccccc|c}
\toprule
Name                & MF      & $n$  & $N_\uparrow + N_\downarrow$      & $K$ & HF Energy       & CCSD            & Ours & FCI     \\
\midrule
Hydrogen            & H2      & 4     & 1+1=2     & 15      & -1.06610864     & \textbf{-1.101150}   & \textbf{-1.101150}    & -1.101150 \\
Lithium Hydride     & LiH     & 12    & 2+2=4     & 631     & -7.76736213     & -7.784455       & \textbf{-7.784460}    & -7.784460 \\
Water               & H2O     & 14    & 5+5=10    & 1390    & -74.9644475     & -75.015409      & \textbf{-75.015511}   & -75.015530 \\
Methylene           & CH2     & 14    & 5+3=8      & 2058    & -37.4846329     & -37.504411      & \textbf{-37.504419}   & -37.504435 \\
Beryllium Hydride   & BeH2    & 14    & 3+3=6     & 2074    & -14.4432411     & -14.472713      & \textbf{-14.472922}   & -14.472947 \\
Ammonia             & NH3     & 16    & 5+5=10    & 4929    & -55.4547926     & -55.520931      & \textbf{-55.521037}     & -55.521150 \\
Methane             & CH4     & 18    & 5+5=10    & 8480    & -39.7265817     & -39.806022      & \textbf{-39.806170}     & -39.806259 \\
Diatomic Carbon     & C2      & 20    & 6+6=12     & 2239    & -74.2483215     & -74.484727      & \textbf{-74.486037}    & -74.496388 \\
Fluorine            & F2      & 20    & 9+9=18    & 2951    & -195.638041     & \textbf{-195.661086}     & -195.661067             & -195.66108 \\
Nitrogen            & N2      & 20    & 7+7=14    & 2239    & -107.498967     & -107.656080     & \textbf{-107.656763}    & -107.660206 \\
Oxygen              & O2      & 20    & 9+7=16    & 2879    & -147.631948     & -147.747738     & -147.749953    & -147.750235 \\
Lithium Fluoride    & LiF     & 20    & 6+6=12    & 5849    & -105.113709     & -105.159235     & \textbf{-105.165270}    & -105.166172 \\
Hydrochloric Acid   & HCl     & 20    & 9+9=18    & 5851    & -455.135968     & \textbf{-455.156189}     & \textbf{-455.156189}           & -455.156189 \\
Hydrogen Sulfide    & H2S     & 22    & 9+9=18    & 9558    & -394.311379     & -394.354556     & \textbf{-394.354592}    & -394.354623 \\
Formaldehyde        & CH2O    & 24    & 8+8=16    & 20397   & -112.354197     & -112.498567     & \textbf{-112.500944}    & -112.501253 \\
Phosphine           & PH3     & 24    & 9+9=18    & 24369   & -338.634114     & -338.698165     & \textbf{-338.698186}    & -338.698400 \\
Lithium Chloride    & LiCL    & 28    & 10+10=20  & 24255   & -460.827258     & -460.847580     & \textbf{-460.848109}    & -460.849618 \\
Methanol            & CH4O    & 28    & 9+9=18    & 52887   & -113.547027     & \textbf{-113.665485}     & \textbf{-113.665485}    & -113.666485 \\
Lithium Oxide       & Li2O    & 30    & 7+7=14    & 20558   & -87.7955672     & -87.885514      & \textbf{-87.885637}     & - \\
Ethylene Oxide      & C2H4O   & 38    & 12+12=24  & 137218  & -150.927608     & -151.120474     & \textbf{-151.120486}   & - \\
Propene             & C3H6    & 42    & 12+12=24  & 161620  & -115.657941     & -115.885123     & \textbf{-115.886571}   & - \\
Acetic Acid         & C2H4O2  & 48    & 16+16=32  & 461313  & -224.805400     & \textbf{-225.050896}     & -225.0429767            & - \\
Sulfuric Acid       & H2O4S   & 62    & 25+25=50  & 1235816 & -689.262656     & -689.498410     & \textbf{-689.505237}   & - \\
Sodium Carbonate    & CNa2O3  & 76    & 26+26=52  & 1625991 & -575.016102     & -575.299810     & \textbf{-575.299820}   & - \\
\toprule
\end{tabular}
\caption{\small Best molecular ground-state energies obtained by different methods as described in the main text over five trials. Molecules have been sorted according to the number of qubits used in the Jordan-Wigner representation. In addition, the numbers $(N_\uparrow,N_\downarrow)$ up- and down-spin electrons and the number $K$ of terms in the Hamiltonian are reported. Our method exhibits superior performance in comparison with classical approximate methods such as Hartree-Fock and CCSD and comes close to the FCI ground truth, which is only available up to molecules of size 28 qubits.}
\label{tbl:chemistry_performance}
\end{table}

\section{Experiments}\label{sec:experiments}

We now investigate the performance of our algorithm and the running time efficiency of the proposed parallelization paradigm.
We first demonstrate our main results by comparing our algorithm with Hartree-Fock (HF) and CC with up to double excitations (CCSD) baselines, where our performance is either on par or superior to the classical approximate methods over a wide list of chemical molecules. Our performance is also close to the ground truth FCI energies up to molecular systems with 28 qubits, where the results for larger molecules become increasingly hard to obtain. We then perform ablation studies to examine our algorithm over various aspects. First, we show that increasing batch size improves the performance, at the cost of increased algorithmic complexity. On the other hand, our parallelization strategy can effectively reduce the running time, achieving near-optimal weak scaling. Our proposed sampling trick also improves the performance of our model by a noticeable margin. At last, we show that our model architecture exhibits superior running time efficiency compared against RBM \cite{choo-20fermionic} and NADE \cite{barrett2022autoregressive}.

\subsection{Experimental Set-ups}

Given a target molecule ID, we fetch its corresponding \textit{PubChem Compound Identifier (CID)} from the official PubChem website~\cite{wang-pub09}. CID is recognized by \texttt{PubChemPy}, a software that provides a way to interact with \textit{PubChem} in Python, allowing depiction and retrieval of chemical properties, such as the geometry of atoms, number of unpaired electrons, total charge, \textit{etc}.  The mapping from second-quantized Hamiltonian to interacting spinning model is done by transforming fermion operators into qubit operators with Jordan-Wigner~\cite{jordan-93} using \texttt{OpenFermion-Psi4} \cite{mcclean-qst20}. Note that these solvers can also be used to estimate the ground state energies including Hartree-Fock, CC with up to double excitations (CCSD), and FCI. The whole data processing pipeline is automatic without further human interference.

We train the model over 10K iterations with Adam optimizer~\cite{kingma-iclr15} by default at a learning rate of $1 \text{\texttimes} 10^{\text{\textminus}3}$ with standard decay rates for the first- and second-moment estimates of $\beta_1 = 0.9$ and $\beta_2 = 0.99$, respectively; no learning rate scheduler is applied. The batch size for the number of unique samples is fixed to be 1024 throughout our experiments unless specified otherwise. For scalability experiments, each GPU is distributed with a constant mini-batch size $mB$, and the effective batch size is $mB \text{\texttimes} L$, where $L$ is the total number of GPUs available.  All experiments in this work use a single set of hyperparameters and identical training procedures, and the results are the best energies obtained across exactly 5 seeds. Throughout the experiments, the timing benchmarks are performed on Tesla V100-SXM2-16GB and Intel(R) Xeon(R) CPU @ 2.30GHz processors on Google Cloud Platform.

\subsection{Performance Benchmark}

\begin{figure}[t]
\centering
\includegraphics[width=0.41\linewidth]{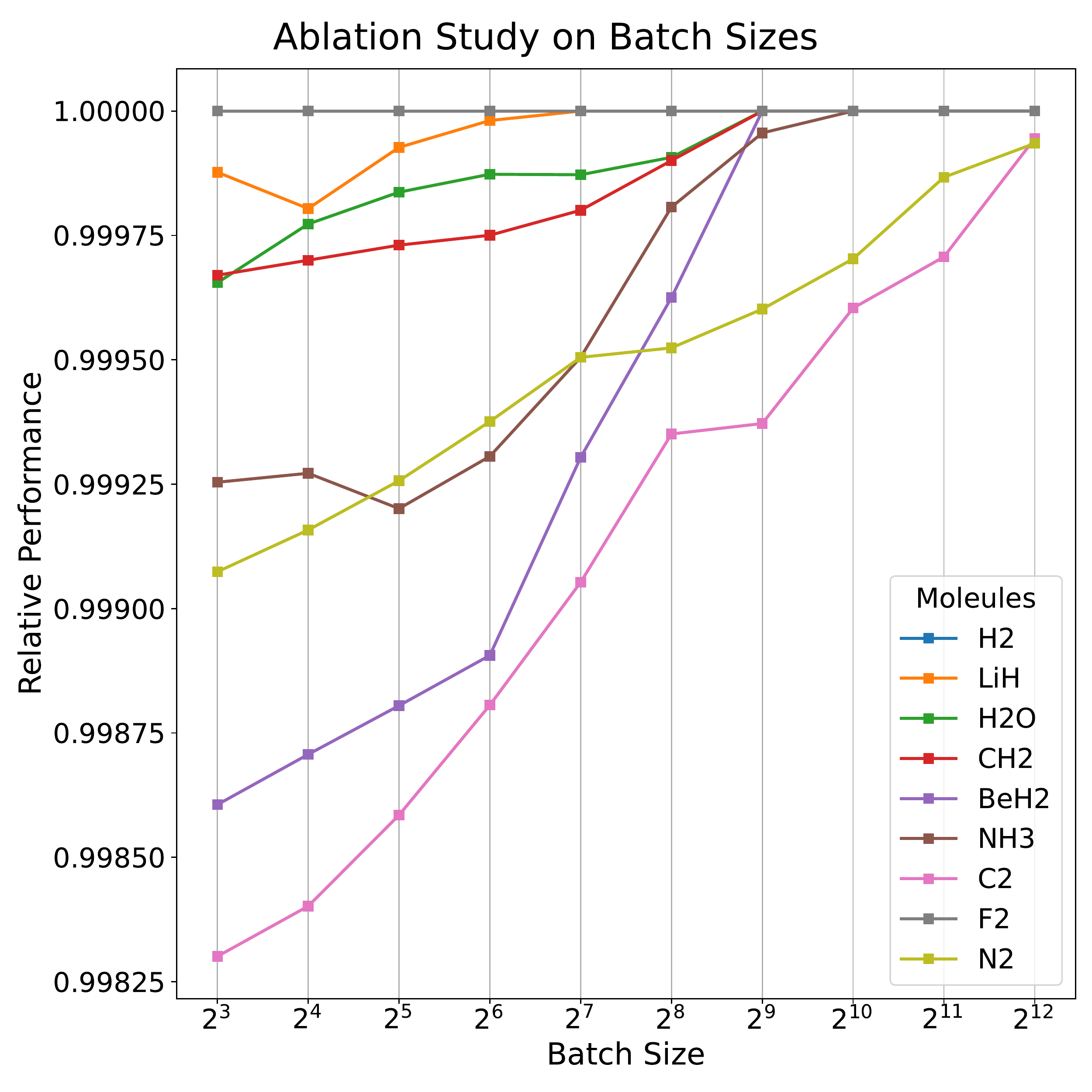}
\includegraphics[width=0.41\linewidth]{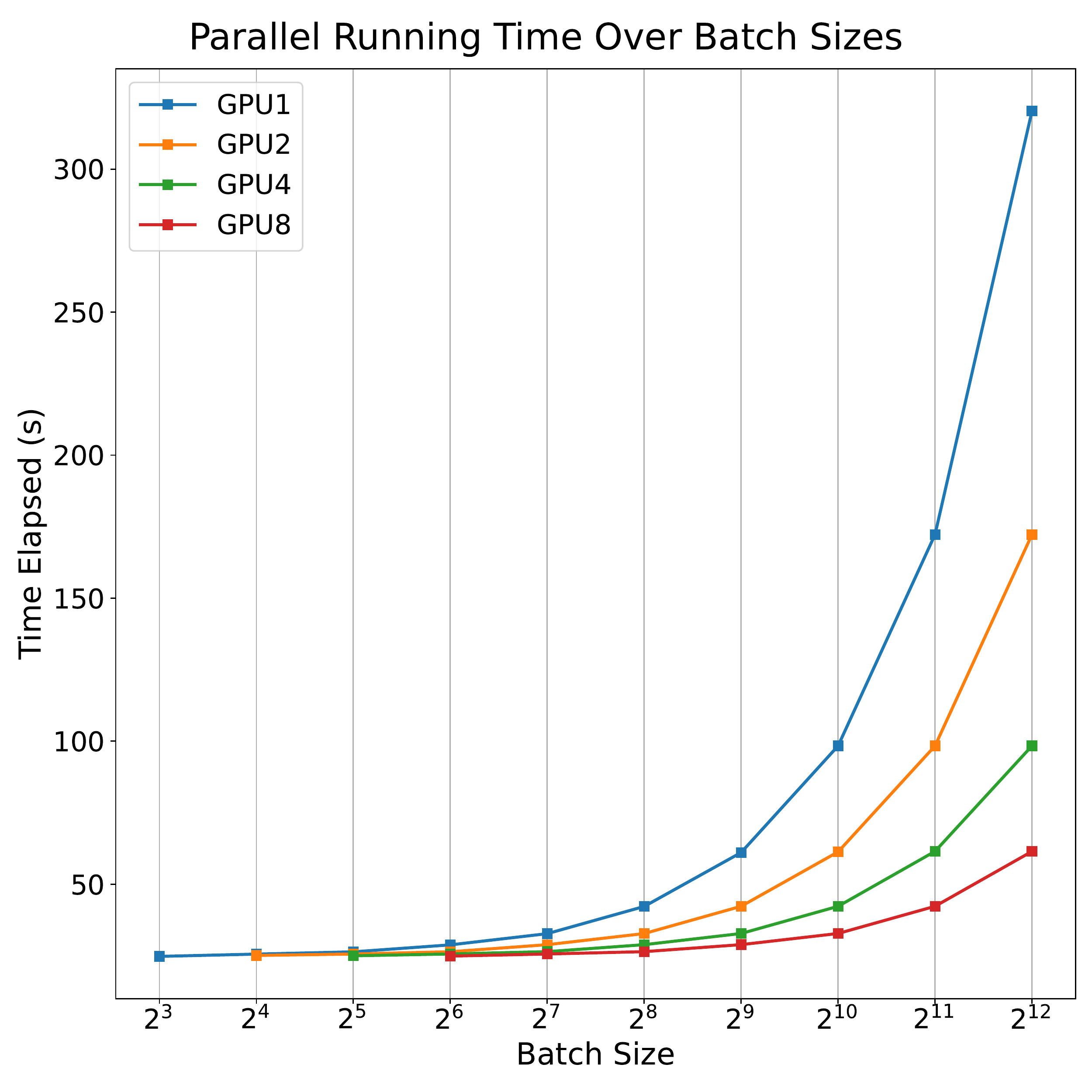}
\caption{Left: The performance of our algorithm with increasing batch size for different molecules. The relative performance score is obtained by normalizing the estimated energy with respect to the corresponding FCI ground truth. Better performance is obtained by training with larger batch sizes, and the improvements become substantial for larger molecules. Right: Demonstration of near-optimal weak scaling obtained by running the algorithm on the C2 molecule for $10^3$ iterations with different batch sizes up to $4096$ and reporting the time elapsed in seconds. Upon distributing the batch over multiple GPUs, the running time is significantly reduced. In addition, the running times for the training with a fixed batch size per GPU are close across different settings.}
\label{fig:batch_size}
\end{figure}

We report the performance of our implementation over a wide variety of molecules in Table \ref{tbl:chemistry_performance}. The molecules are sorted according to the number of qubits in their Jordan-Wigner representation. 
We consider Hartree-Fock (HF) and CCSD energies as classical baseline approximate methods to compare against. The FCI ground truth energy is also provided for smaller molecules for reference. 

Our model exhibits consistently strong performance on all molecules considered. In particular, the computed energies match the ground-truth FCI result closely on all molecules with up to 20 electrons and 28 spin-orbitals and out-perform other approximate methods on the majority of the molecules we considered. Notice that as the size of the molecule increases, the number of terms in the electronic Hamiltonian formulation also grows quickly, which leads to severe computational issues such as running time and memory consumption.  As a result, our proposed algorithm is capable of scaling up to system size with 76 qubits of which the electronic Hamiltonian is consisted of 1.6 million terms, and achieves state-of-the-art performance.

\subsection{Ablation Studies}

In this section, we perform ablation studies to test the effectiveness of the ingredients in our contribution. We start by validating the fact that increasing batch size is capable of improving the performance for larger-scale problems, which justifies the motivation of our parallelization scheme that enables large-batch training for large molecules. We also examine our proposed reverse sampling trick by comparing the performance with different sampling orders. Finally, we tried different model architectures under the same training framework; our model achieves the overall best running time efficiency in comparison with RBM \cite{choo-20fermionic} and NADE \cite{barrett2022autoregressive}.

\xhdr{Parallelization.} As discussed in Section~\ref{sec:formulation}, the total number of input samples for the forward pass required to compute the local energy scales with the number of terms $K$ in the Hamiltonian and with the batch size $B$, which leads to a heavy computational bottleneck as both of these factors increase.
However, a sufficiently large batch size is essential to guarantee the performance of the algorithm for molecules of larger sizes.
This claim is validated in the left panel of Figure~\ref{fig:batch_size}, where we train our model with batches of varying sizes for various molecules. Since the ground state energies for different molecules differ from each other, we report the performance relative the FCI ground truth.

To examine the effectiveness of our parallelization scheme, in the right panel of Figure~\ref{fig:batch_size} we illustrate the running time as a function of batch size for C2 molecule using $10^3$ iterations. We observe that the running time scales inversely with respect to the number of GPUs. In particular, doubling the number of GPUs roughly corresponds to half the time usage. We conducted additional experiments by saturating the memory on each GPU and observe that the running time for different numbers of GPUs remains constant, indicating that our approach achieves near-optimal weak scaling.

\xhdr{Reverse Order Sampling.} We proposed to perform autoregressive sampling in a reversed order to improve the training. To examine the effectiveness of this approach as well as the impact of the sampling order on quantum chemistry problems in general, we perform ablation studies on the sampling order. In Table~\ref{tbl:sample_order}, we employ three different sampling orders: forward sampling from 1 to $n$, reverse sampling from $n$ to 1, and random sampling by any pre-determined order between 1 to $n$. The results illustrate that reverse sampling indeed improves the performance effectively as it achieves the best results consistently across the list of molecules.

\begin{table}[t]
\scriptsize
\centering
\begin{tabular}{l|cccccccc}
\toprule
Molecule & H2 & H2O & NH3 & C2 & N2 & O2 & HCL \\
\midrule
Energy & \multicolumn{7}{c}{Ablation study on the sampling order} \\
\midrule
Forward      & \textbf{-1.101150} & -75.015449 & -55.515394 & -74.4849249 & -107.634908 & -147.723681 & -454.927860 \\
Reverse      & \textbf{-1.101150} & \textbf{-75.015511} & \textbf{-55.521037} & \textbf{-74.4860377} & \textbf{-107.656763} & \textbf{-147.749953} & \textbf{-455.156189} \\
Random       & \textbf{-1.101150} & -75.014553 & -55.519741 & -74.4851979 & -107.606417 & -147.732876 & -455.012498 \\
\midrule
Running time (s) & \multicolumn{7}{c}{Running time for 30K iterations} \\
\midrule
NADE     & 303.76 & 1087.52 & 4474.94 & 4574.30 & 2959.28 & 2821.60 & 1593.21 \\
MADE     & \textbf{282.64} & \textbf{838.60} & \textbf{3188.31} & \textbf{2922.25} & \textbf{2295.90} & \textbf{2122.21} & \textbf{1112.70} \\
\midrule
Hitting time (s) & \multicolumn{7}{c}{Hitting time to the CCSD performance} \\
\midrule
NADE     & 117.28 & \textbf{352.78} & 2007.14 & 1754.10 & 1029.64 & 824.37 & 489.88 \\
MADE     & \textbf{100.14} & 364.32 & \textbf{782.46} & \textbf{827.41} & \textbf{986.27} & \textbf{648.27} & \textbf{186.38} \\
\toprule
\end{tabular}
\caption{\small Ablation study on reverse sampling and time efficiency tests over different architectures.}
\label{tbl:sample_order}
\end{table}

\begin{figure}[t]
\centering
\includegraphics[width=0.80\linewidth]{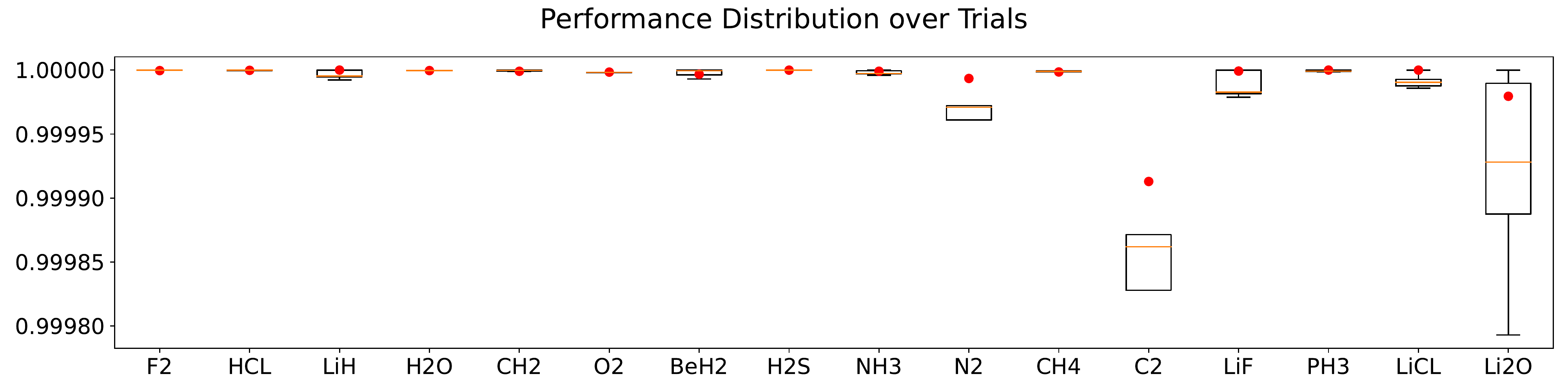}
\caption{We report our performance on different molecules for 5 trials in the form of a box plot. For illustration purposes, we divide the results for all molecules by the corresponding FCI ground truths, so that the reported value is normalized with a maximum value of 1. In addition, we directly cite the numbers in Table 1 from \cite{barrett2022autoregressive} and mark them in the form of red dots in the figure for comparison purposes. We notice that for certain molecules, the performance of our method fluctuates due to different random seeds. Nevertheless, competitive results can be obtained after a sufficient number of trials.}
\label{fig:compare}
\end{figure}

\xhdr{Model Architecture.} Our model architecture offers significant parallelization advantages compared to existing architectures based on RBM and NADE. Despite the sequential nature of both MCMC and autoregressive sampling, the latter can be executed on GPUs in a straightforward fashion and moreover exhibits superior running time efficiency. In addition, the distribution of the MCMC samples only converges to the distribution of interest asymptotically, whereas autoregressive sampling yields exact samples under a known number of iterations. NADE~\cite{larochelle2011neural} requires $n$ forward passes
through the network to evaluate the probability, with $n$ submodules for each entry. The main disadvantage of NADE is its sequential nature in its forward pass, which contributes to running time as the input dimension grows. In addition, computation with NADE is slower in practice compared to MADE even in low dimensions, especially for the model of high depth, due to its multi-module architectural design. In Table~\ref{tbl:sample_order}, we directly compare the time taken for NADE and MADE to run for $3\times10^4$ iterations. In addition, we measure the time taken for each architecture to reach the performance of CCSD for each molecule. We did not include the results for RBM for two reasons. First, the actual running times of RBM for 30K iterations are exceedingly large, \textit{e.g.}, about four hours for H2 molecule, therefore the direct comparison results against the other two architectures have no practical interest. Second, the performance of RBM is inferior to CCSD, so the hitting time is not available. On the other hand, MADE exhibits a significant computational advantage compared to NADE in both measures. We also directly cite the numbers from Table 1 in \cite{barrett2022autoregressive} and compare the results with ours in Figure~\ref{fig:compare} for a sanity check.

\section{Conclusions}

We proposed a scalable parallelization strategy to improve the VMC algorithm in the application of ab-initio quantum chemistry. Our local energy parallelism enables the optimization for Hamiltonians of more complex molecules and our autoregressive sampling techniques out-perform the CCSD baseline and exhibit an advantage against other neural-network based algorithms in terms of running time efficiency and scalability. We further improve the performance of our model through the sampling order of the state entries to match the entanglement hierarchy among the molecule qubits. Our algorithm effectively works for molecules up to 76 qubits with millions of terms in its electric Hamiltonian.






\bibliographystyle{unsrt}  
\bibliography{references}

\begin{thebibliography}{10}

\bibitem{troyer-pr05}
Matthias Troyer and Uwe-Jens Wiese.
\newblock Computational complexity and fundamental limitations to fermionic
  quantum monte carlo simulations.
\newblock {\em Physical review letters}, 94(17):170201, 2005.

\bibitem{coester-60short}
Fritz Coester and Hermann K{\"u}mmel.
\newblock Short-range correlations in nuclear wave functions.
\newblock {\em Nuclear Physics}, 17:477--485, 1960.

\bibitem{bartlett-07coupled}
Rodney~J Bartlett and Monika Musia{\l}.
\newblock Coupled-cluster theory in quantum chemistry.
\newblock {\em Reviews of Modern Physics}, 79(1):291, 2007.

\bibitem{choo-20fermionic}
Kenny Choo, Antonio Mezzacapo, and Giuseppe Carleo.
\newblock Fermionic neural-network states for ab-initio electronic structure.
\newblock {\em Nature communications}, 11(1):1--7, 2020.

\bibitem{barrett2022autoregressive}
Thomas~D Barrett, Aleksei Malyshev, and AI~Lvovsky.
\newblock Autoregressive neural-network wavefunctions for ab initio quantum
  chemistry.
\newblock {\em Nature Machine Intelligence}, 4(4):351--358, 2022.

\bibitem{germain-icml15}
Mathieu Germain, Karol Gregor, Iain Murray, and Hugo Larochelle.
\newblock Made: Masked autoencoder for distribution estimation.
\newblock In {\em International Conference on Machine Learning}, pages
  881--889, 2015.

\bibitem{zhao2021overcoming}
Tianchen Zhao, James Stokes, Oliver Knitter, Brian Chen, and Shravan
  Veerapaneni.
\newblock Overcoming barriers to scalability in variational quantum monte
  carlo.
\newblock {\em The International Conference for High Performance Computing,
  Networking, Storage, and Analysis}, 2021.

\bibitem{salakhutdinov2008quantitative}
Ruslan Salakhutdinov and Iain Murray.
\newblock On the quantitative analysis of deep belief networks.
\newblock In {\em International Conference on Machine Learning}, pages
  872--879, 2008.

\bibitem{larochelle2011neural}
Hugo Larochelle and Iain Murray.
\newblock The neural autoregressive distribution estimator.
\newblock In {\em Proceedings of the Fourteenth International Conference on
  Artificial Intelligence and Statistics}, pages 29--37, 2011.

\bibitem{carleo2017solving}
Giuseppe Carleo and Matthias Troyer.
\newblock Solving the quantum many-body problem with artificial neural
  networks.
\newblock {\em Science}, 355(6325):602--606, 2017.

\bibitem{bengio2000modeling}
Yoshua Bengio and Samy Bengio.
\newblock Modeling high-dimensional discrete data with multi-layer neural
  networks.
\newblock {\em Advances in Neural Information Processing Systems}, 12:400--406,
  2000.

\bibitem{sharir-prl20}
Or~Sharir, Yoav Levine, Noam Wies, Giuseppe Carleo, and Amnon Shashua.
\newblock Deep autoregressive models for the efficient variational simulation
  of many-body quantum systems.
\newblock {\em Physical review letters}, 124(2):020503, 2020.

\bibitem{sharir-git20}
Or~Sharir, Yoav Levine, Noam Wies, Giuseppe Carleo, and Amnon Shashua.
\newblock Flowket: an open-source library based on tensorflow for running
  variational monte-carlo simulations on gpus.
\newblock \url{https://github.com/HUJI-Deep/FlowKet}, 2020.

\bibitem{hibat2020recurrent}
Mohamed Hibat-Allah, Martin Ganahl, Lauren~E Hayward, Roger~G Melko, and Juan
  Carrasquilla.
\newblock Recurrent neural network wave functions.
\newblock {\em Physical Review Research}, 2(2):023358, 2020.

\bibitem{oord2016pixelcnn}
Aaron van~den Oord, Nal Kalchbrenner, Lasse Espeholt, koray kavukcuoglu, Oriol
  Vinyals, and Alex Graves.
\newblock Conditional image generation with pixelcnn decoders.
\newblock In D.~Lee, M.~Sugiyama, U.~Luxburg, I.~Guyon, and R.~Garnett,
  editors, {\em Advances in Neural Information Processing Systems}, volume~29.
  Curran Associates, Inc., 2016.

\bibitem{hammond-94monte}
Brian~L Hammond, William~A Lester, and Peter~James Reynolds.
\newblock {\em Monte Carlo methods in ab initio quantum chemistry}, volume~1.
\newblock World Scientific, 1994.

\bibitem{langhoff-12quantum}
S~Langhoff.
\newblock {\em Quantum mechanical electronic structure calculations with
  chemical accuracy}, volume~13.
\newblock Springer Science \& Business Media, 2012.

\bibitem{sherrill-99advances}
C~David Sherrill and HF~Schaefer~III.
\newblock Advances in quantum chemistry.
\newblock {\em Advances in Quantum Chemistry}, 34:143--269, 1999.

\bibitem{born1927quantentheorie}
Max Born and Robert Oppenheimer.
\newblock Zur quantentheorie der molekeln.
\newblock {\em Annalen der physik}, 389(20):457--484, 1927.

\bibitem{jordan-93}
Pascual Jordan and Eugene~Paul Wigner.
\newblock {\"u}ber das paulische {\"a}quivalenzverbot.
\newblock In {\em The Collected Works of Eugene Paul Wigner}, pages 109--129.
  1993.

\bibitem{bravyi-02fermionic}
Sergey~B Bravyi and Alexei~Yu Kitaev.
\newblock Fermionic quantum computation.
\newblock {\em Annals of Physics}, 298(1):210--226, 2002.

\bibitem{lin2017deep}
Yujun Lin, Song Han, Huizi Mao, Yu~Wang, and William~J Dally.
\newblock Deep gradient compression: Reducing the communication bandwidth for
  distributed training.
\newblock In {\em International Conference on Learning Representations}, 2017.

\bibitem{wang-pub09}
Yanli Wang, Jewen Xiao, Tugba~O Suzek, Jian Zhang, Jiyao Wang, and Stephen~H
  Bryant.
\newblock Pubchem: a public information system for analyzing bioactivities of
  small molecules.
\newblock {\em Nucleic acids research}, 37:W623--W633, 2009.

\bibitem{mcclean-qst20}
Jarrod~R McClean, Nicholas~C Rubin, Kevin~J Sung, Ian~D Kivlichan, Xavier
  Bonet-Monroig, Yudong Cao, Chengyu Dai, E~Schuyler Fried, Craig Gidney,
  Brendan Gimby, et~al.
\newblock Openfermion: the electronic structure package for quantum computers.
\newblock {\em Quantum Science and Technology}, 5(3):034014, 2020.

\bibitem{kingma-iclr15}
Diederik~P. Kingma and Jimmy Ba.
\newblock Adam: {A} method for stochastic optimization.
\newblock In {\em International Conference on Learning Representations}, 2015.

\end{thebibliography}

\end{document}